\documentclass[referee]{raa}
\usepackage{graphicx,times}
\usepackage{natbib}
\usepackage{amssymb,amsmath}

\begin{document}

   \title{A new statistical distance scale for planetary nebulae, based on Gaia EDR3
}

 \volnopage{ {\bf 20XX} Vol.\ {\bf X} No. {\bf XX}, 000--000}
   \setcounter{page}{1}

   \author{Ali, A.\inst{1}$^*$ \footnotetext{\small $*$ Corresponding author}, Algarni, E.\inst{2}, Mindil, A.\inst{3}, Alghamdi, S.A.\inst{2}}


   \institute{Astronomy, Space Science \& Meteorology Department, Faculty of Science, Cairo University, Giza 12613, Egypt.; {\it afouad@sci.cu.edu.eg}\\
        \and
             Astronomy and space science department, Faculty of Science, King Abdulaziz University, Jeddah, Saudi Arabia.\\
	\and
Department of Physics, College of Science, University of Jeddah, Jeddah, Saudi Arabia.; {\it amindil@uj.edu.sa}\\
\vs \no
   {\small Received 20XX Month Day; accepted 20XX Month Day}
}

\abstract{The present work aims to build a new statistical distance scale for planetary nebulae (PNe) based on a rigorous calibration sample. The distances of the calibration sample are derived from the trigonometric parallax method using the recent measurements of Gaia’s early third data release (Gaia EDR3). The new distance scale is created by applying the well-known linear relationship between the radio surface brightness temperature and the nebular radius. The calibration sample is made up of 96 PNe of accurately computed distances with uncertainties less than $20\%$. Earlier ground- and space-based trigonometric parallaxes of PNe display inconsistency with those of Gaia, particularly the HIPPARCOS results.  In addition, these measurements have appreciably lower precision than that of Gaia. When compared to the trigonometric technique, the expansion and kinematic methods exhibited more consistency than the spectroscopic, extinction, gravity, and photo-ionization methods. Furthermore, contrary to earlier results in the literature, the extinction and gravity methods, on average, underestimate and slightly overestimate the PN distances.  As a byproduct of extracting the Gaia parallaxes, we detect the radial velocity and variability for 14 and 3 PN central stars (CSs), respectively. To our knowledge, the variability of Hen 2-447 CS has been determined for the first time.
}

   \authorrunning{Ali et al. }            
   \titlerunning{A new statistical distance scale for PNe}  
   \maketitle

%
\section{Introduction}           
\label{sect:intro}

Gaia is a space mission that was launched and operated by the European Space Agency (ESA) to provide a detailed three-dimensional map of the Milky Way galaxy. The first Gaia data release (Gaia DR1) was published in September 2016, followed by the second Gaia data release (Gaia DR2) in April 2018 and the early third data release (Gaia EDR3) in December 2020. The full Gaia data release (Gaia DR3) is scheduled in the first part of 2022. Gaia EDR3 provides the position and apparent magnitude for $\sim 1.8$ billion sources, as well as the parallax ($\pi_{\rm Gaia}$), proper motion ($\mu$), and (B-R) color for $\sim 1.5$ billion sources. In comparison to the Gaia DR2, the newest release  exhibits considerable enhancements in the astrometric and photometric accuracy, precision, and homogeneity (Gaia Collaboration et al., 2020). The precision of the parallax and proper motion are improved by $30\%$ and a factor of two, respectively. Moreover, the estimated parallax zero-point for Gaia EDR3 data, $-0.017$ mas \citep{Lindegren21}, was enhanced compared with the Gaia DR2, $-0.029$ mas \citep{Lindegren18}.

The precise distances of PNe enable astronomers to better understand the evolution of low and intermediate-mass stars and the entire Galaxy. Knowing the distance is a key tool for studying the significant parameters of the PN and its accompanying CS. However, because of the wide variety of their characteristics, obtaining accurate distances for PNe is a difficult task. The procedures usually applied to derive the distances of PNe are known as individual and statistical methods. The description, limitations, assumptions, and uncertainties of these methods were discussed in \citet{Frew16}, hereafter FBP16. Although the trigonometric technique is the only direct and trusted individual method for defining PN distances, however, it is confined to nearby PNe that are linked with detected CSs. Therefore, there is still a need to apply other methods to determine the distances of remote PNe and those associated with very faint or undetectable CSs.

Any statistical method depends on a relationship between two nebular parameters, one is distance-dependent and the other is distance-independent. After calibrating such a relationship using PNe of known distances, we can use it to calculate the statistical distance to any PN.
\citet{Ali15}, hereafter AIA15, have developed two statistical distance scales based on a calibration sample composed of 82 PNe, which is larger and more dependable than that applied in prior scales at that time. Except for a few objects with trustworthy trigonometric, spectroscopic, and cluster membership distances, the distance of each calibrator was computed as a weighted mean value derived from at least two different individual methods. This sample was applied to recalibrate the linear mass-radius ($M-R$) and radio surface brightness temperature-radius ($Tb-R$) relationships. The main goal of this study is to improve the $Tb-R$ distance scale by using a more precise, reliable, and larger calibration sample than that used by AIA15.

\citet{Stanghellini2020} proposed a PN statistical distance scale, based on the linear relationship between the nebular radius and $H_{\beta}$ surface brightness. This scale was calibrated using a PN sample of distances extracted from the Gaia DR2 parallaxes.  This distance scale is defined as: ${\rm log}(R) = -(0.226 \pm 0.0155) \times {\rm log}(S_{Hb}) - (3.920 \pm 0.215)$, where $R$ is the nebular radius in pc and $S_{Hb}$ is the $H_{\beta}$ surface brightness.

Recently, \citet{Chornay21} have released a catalogue of 2118 CSs from the Gaia EDR3. Examining the catalogue, we found 424 and 351 PNe with unknown and  negative parallaxes, respectively. From the remaining list, there are 67, 361, and 915 CSs of unknown, blue, and red colors, respectively.

The objectives of this article can be summarised as follows: (1) updating the $Tb-R$ distance scale presented in AIA15, applying a more accurate, homogenous, and reliable calibration sample of distances taken from Gaia EDR3; (2) comparing the PN Gaia parallaxes to past measurements in the literature; (3) examining the consistency between the trigonometric and other individual distance methods, based on larger statistical samples than presented in preceding studies. As a byproduct of this study, we detect the radial velocity of 14 PNe and the stellar variability of three PN CSs.

Sections 2 and 3 discuss the calibration sample as well as comparisons between Gaia and earlier parallax measurements.  Section 4 discusses the consistency between the trigonometric and other individual distance methods, whereas Section 5 presents the new distance scale. Section 6 displays the identification of 14 CS radial velocities and stellar variability of three PN CSs, while the conclusion is given in Section 7.

\begin{table}
\centering
\caption{Gaia EDR3 sources that matched true, possible, and likely PNe in the HASH catalogue.} \label{Table1}
\scalebox{0.6}{
\begin{tabular}{llccccccccccc}
\hline
PN name	&	Gaia  designation	&	l	&	b	&	$\alpha$	&	$\delta$	&	$\pi_{\rm Gaia}$	&	$\mu_{\alpha}$ &	$\mu_{\delta}$ 	&	G\,mag	&	B\,mag	&	R\,mag	&	B-R	\\
\hline
PN PC 12	&	Gaia EDR3 4130784921205604736	&	0.1660	&	17.2488	&	250.9741	&	-18.9533	&	0.052	$\pm$	0.066	&	-3.17	$\pm$	0.08	&	-2.37	$\pm$	0.06	&	15.2	&	14.0	&	13.4	&	0.62	\\		
IC 4634	&	Gaia EDR3 4126115570219432448	&	0.3619	&	12.2147	&	255.3899	&	-21.8259	&	0.353	$\pm$	0.043	&	-1.20	$\pm$	0.05	&	-5.12	$\pm$	0.03	&	13.9	&	12.3	&	12.4	&	-0.15	\\		
PN G000.7+08.0	&	Gaia EDR3 4114088875802922880	&	0.7176	&	8.0618	&	259.2877	&	-23.9416	&	0.045	$\pm$	0.254	&	-3.39	$\pm$	0.29	&	-5.42	$\pm$	0.20	&	18.9	&	19.2	&	18.5	&	0.74	\\		
PN Bl 3-13	&	Gaia EDR3 4056540677880158208	&	0.9497	&	-2.0864	&	269.0116	&	-29.1880	&	0.150	$\pm$	0.183	&	-1.96	$\pm$	0.18	&	-5.51	$\pm$	0.13	&	18.0	&	16.7	&	15.5	&	1.15	\\		
PN G001.2-05.6	&	Gaia EDR3 4049240298544263936	&	1.2146	&	-5.6614	&	272.7613	&	-30.7033	&	0.650	$\pm$	0.422	&	-2.86	$\pm$	0.44	&	-0.39	$\pm$	0.35	&	18.6	&	18.5	&	18.2	&	0.35	\\		
PN H 1-47	&	Gaia EDR3 4062301564840251520	&	1.2949	&	-3.0402	&	270.1568	&	-29.3641	&	0.074	$\pm$	0.052	&	1.32	$\pm$	0.06	&	-7.02	$\pm$	0.04	&	15.7	&	15.2	&	13.9	&	1.26	\\		
PN SwSt 1	&	Gaia EDR3 4049331244394134912	&	1.5906	&	-6.7176	&	274.0511	&	-30.8689	&	0.326	$\pm$	0.110	&	-6.24	$\pm$	0.13	&	-1.72	$\pm$	0.09	&	11.8	&	11.0	&	10.1	&	0.85	\\		
PN H 1-55	&	Gaia EDR3 4050131349653595392	&	1.7136	&	-4.4554	&	271.8107	&	-29.6902	&	0.019	$\pm$	0.860	&	-2.44	$\pm$	0.78	&	1.41	$\pm$	0.57	&	16.6	&	15.4	&	14.3	&	1.13	\\		
PN H 1-56	&	Gaia EDR3 4050126711087206016	&	1.7359	&	-4.6055	&	271.9745	&	-29.7429	&	0.075	$\pm$	0.062	&	-4.00	$\pm$	0.07	&	-9.42	$\pm$	0.05	&	16.0	&	14.7	&	14.3	&	0.38	\\		
PN M 2-33	&	Gaia EDR3 4049596604655713664	&	2.0229	&	-6.2249	&	273.7773	&	-30.2593	&	0.224	$\pm$	0.042	&	-0.06	$\pm$	0.04	&	-6.03	$\pm$	0.03	&	14.7	&	13.9	&	13.8	&	0.14	\\		

\hline
\end{tabular}}
\end{table}

\section{The calibration sample} \label{The calibration sample}
From $\sim 3500$ known Galactic PNe \citep{Parker12}, there are 620 CSs that are spectroscopically confirmed as single/binary PN ionizing stars in the recent catalogue of \citet{Weidmann20}. The catalogue includes some misclassified objects such as EGB 4 (Nova-like Star), K 2-15 (HII region), WRAY 16-193 (symbiotic Star), and LS III +51 42 (emission-line star). The intense ultraviolet radiation output of most PN CSs causes them to appear as blue stars, however, there are many CSs that appear as red stars. This can be attributed to either the high reddening of its line of sight or the visible light being dominated by its close binary main sequence companion. Although the CS usually lies at the geometric centre of the nebula, more evolved PNe and those interacting with the interstellar medium (ISM) have shown off-center shifts.

We restrict our search in the Gaia EDR3 database to the blue CSs of PNe that are listed in the HASH catalogue \citep{Parker16} as true, possible, and likely PNe.  In addition, we complement our sample with red CSs that have been spectroscopically confirmed as PN nuclei by \citet{Weidmann20}. Further, we reject all matched Gaia sources of negative and missed parallax as well as those of unknown colors. We collect 603 matched Gaia sources. A part of this data is given in Table \ref{Table1}, while the full table will be available online. Columns 1, 2, 3-4, and 5-6 provide the object name, Gaia EDR3 designation, equatorial coordinate, and Galactic coordinate, respectively. The parallax and proper motion and their uncertainties are given in columns 6-7 and 8-9, respectively. The stellar magnitudes G, B, R, and the B-R colour index are listed in columns 10, 11, 12, and 13 respectively.

To obtain a high confidence calibrating sample, we select the true PNe from the HASH catalogue with CSs of parallax errors less than $20\%$. The parallax measurements are corrected for the zero-point shift. Further, to recommend the goodness-of-fit indices for the Gaia EDR3 astrometry, we ignored the parallax measurements with a renormalized unit weight error (RUWE) larger than 1.4 \citep{Fabricius21}. Finally, a total of 241 PNe are obtained. Unfortunately, only 95 PNe from this collection have published angular radius ($\theta$) and 5GHz radio surface flux ($F_{5GHz}$). This sample is used to re-calibrate the $Tb-R$ relationship. The farthest PN in the calibration sample is $\sim 6000$ pc away.

To test the new distance scale for calculating distances to remote nebulae, we support the calibration sample with the  nebula "PS1", which belongs to the globular cluster "Pease 1". \citet{McNamara04} calculated a distance of 9.98 kpc for this cluster, which is close to the traditional estimate of 10.4 kpc derived by \citet{Durrell93} but less than the distance of 11.2 kpc that was obtained by \citet{Kraft03}. In this study, we adopted the dynamical distance of $10.3\pm0.4$ kpc that given by \citet{van_den_Bosch06} who developed orbit-based axisymmetric  models for the globular cluster. These models matched 1264 line-of-sight velocity measurements (that extend out to 7 arc-minute) and a sample of 703 proper motions (covering 0.25 arc-minute of the inner cluster part). This enabled them to constrain the change in mass-to-light ratio as a function of radius and calculate the cluster's dynamical distance, inclination, central mass, and density of the cluster.

The full calibration sample will be available online, while a portion is presented in Table \ref{Table2}. The PNG number, PN common name, Gaia EDR3 designation, $F_{5GHz}$ in mJy, $\theta$ in arcseconds, Gaia distance ($D_{G}$) in pc, PN radius ($R$) in pc, and $Tb$ in K are given in columns 1, 2, 3, 4, 5, 6, 7, and 8, respectively.  The predicated distance ($D$) in pc and its residual (see, Section \ref{Tb-R relationship}) are listed in columns 9 and 10, respectively.

\begin{figure}
   \centering
\includegraphics[width=14.0cm, angle=0]{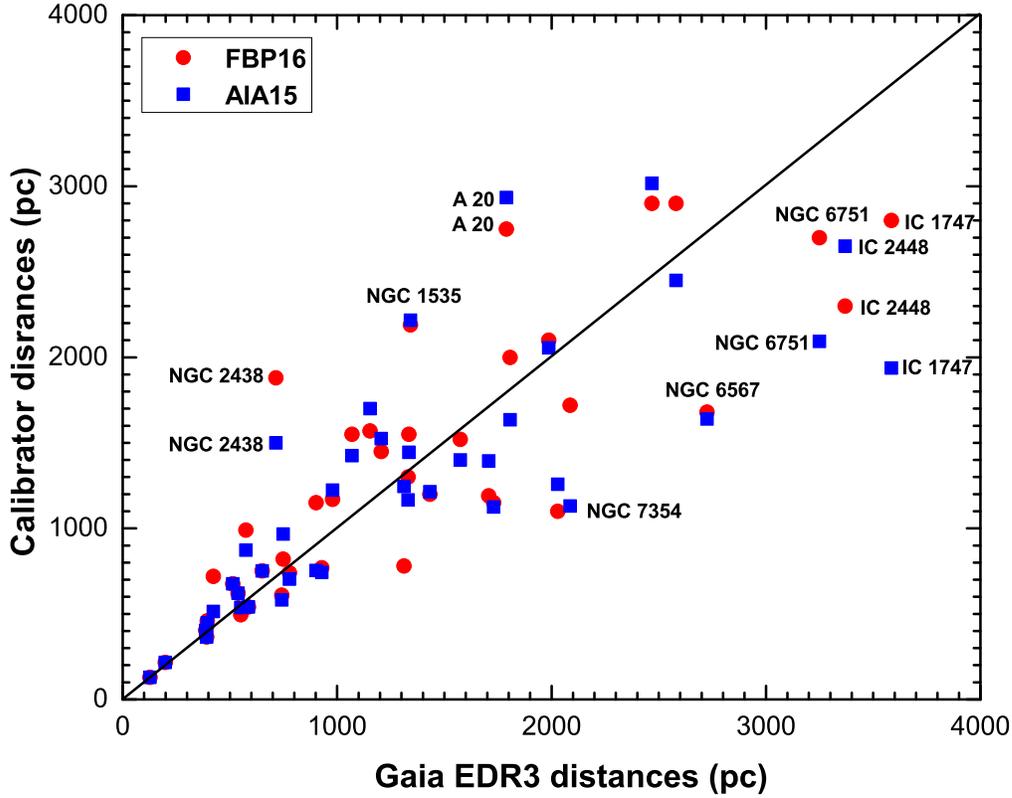}
   \caption{The Gaia EDR3 versus AIA15 and FBP16 calibration samples. }
   \label{Figure1}
   \end{figure}

In Figure \ref{Figure1}, we compare the Gaia EDR3 distances with the calibrator distances adopted by AIA15 and FBP16 statistical scales.  The comparison is based on 41 common objects between both calibration samples and Gaia EDR3. There is a clear match for numerous data points. Despite most of the objects being clustered around the 1:1 line, there are a few exceptions, including IC 1747, NGC 2438, NGC 2438 and A 20. We estimate the median distance ratio between both calibrator samples and Gaia EDR3, where the results reveal that  AIA15  and FBP16 distance scales slightly underestimate (0.98) and overestimate (1.06) the PN Gaia distance, respectively. This comparison is limited to PNe of distance less than $\sim 3600$ pc.

\begin{figure}
{ \begin{tabular}{@{}ccc@{}}
	\includegraphics[scale=0.40]{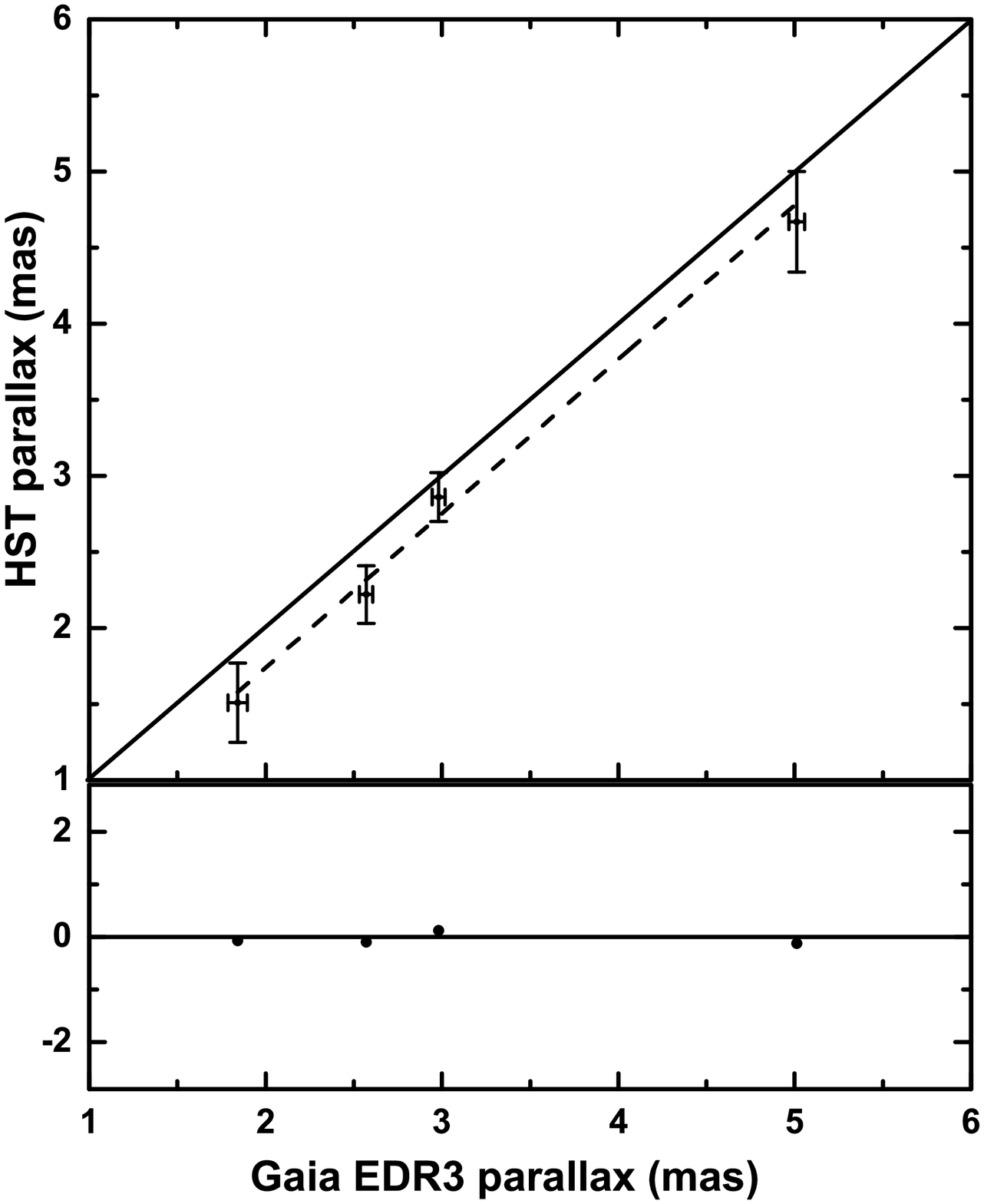} &
     \includegraphics[scale=0.40]{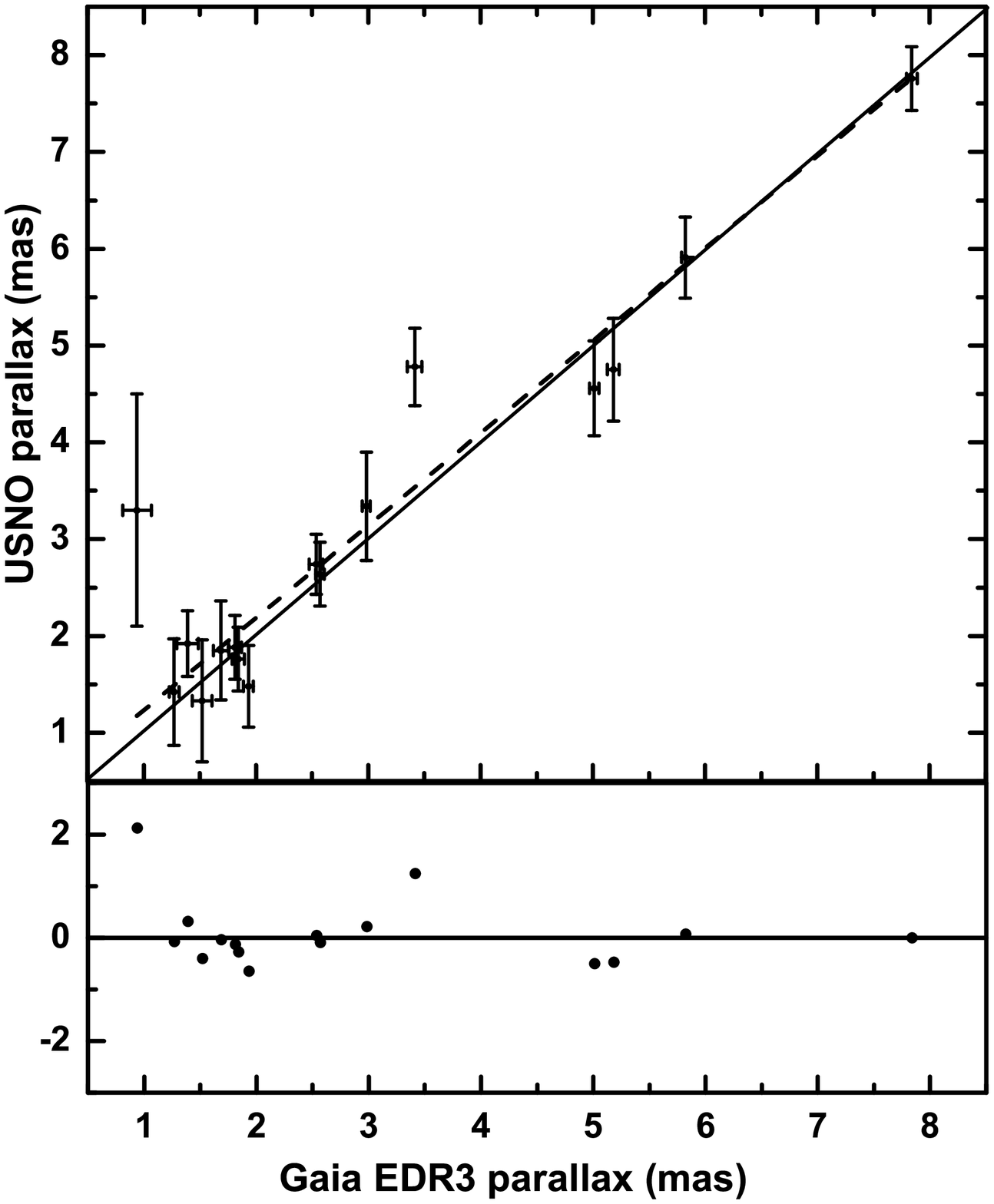}
		\end{tabular}}
   \caption{The Gaia EDR3 against HST (left panel) and USNO (right panel) parallaxes. The solid and dashed lines indicate the 1:1 matches and linear fitting, respectively. The fitting residual is given below each graph.}
   \label{Figure2}
   \end{figure}
   
\begin{table}
\centering
\caption{The calibration sample.} \label{Table2}
\scalebox{0.60}{
\begin{tabular}{lllllllllc}
\hline

PNG 	&	NAME	&	Gaia designation	&	S5GHz (mJy)			&	$\theta$ (")			&	$D_{G}$ (pc)			&	$R$ (pc)			&	$Tb$ (K)			&	$D$ (pc)			&	$|(D_{G}-D)/D_{G}| \%$	\\
\hline
PN G000.3+12.2	&	IC 4634	&	Gaia EDR3 4126115570219432448	&	114.8	$\pm$	16.0	&	5.5	$\pm$	1.10	&	2582	$\pm$	288	&	0.069	$\pm$	0.010	&	69.74	$\pm$	17.00	&	3068	$\pm$	182	&	18.8	\\
PN G002.4+05.8 	&	NGC 6369	&	Gaia EDR3 4111368477921050368	&	2002.0	$\pm$	65.0	&	14.4	$\pm$	2.88	&	1069	$\pm$	56	&	0.075	$\pm$	0.008	&	177.30	$\pm$	35.92	&	912	$\pm$	42	&	14.7	\\
PN G002.7-52.4	&	IC 5148	&	Gaia EDR3 6574225217863069056	&	28.1	$\pm$	3.5	&	62.5	$\pm$	12.51	&	1145	$\pm$	70	&	0.347	$\pm$	0.041	&	0.13	$\pm$	0.03	&	1443	$\pm$	82	&	26.1	\\
PN G009.4-05.0	&	NGC 6629	&	Gaia EDR3 4089517157442187008	&	266.0	$\pm$	53.2	&	8.5	$\pm$	1.69	&	1988	$\pm$	90	&	0.082	$\pm$	0.009	&	67.93	$\pm$	15.19	&	2003	$\pm$	106	&	0.7	\\
PN G009.6+14.8	&	NGC 6309	&	Gaia EDR3 4141505881131938560	&	134.5	$\pm$	26.9	&	8.8	$\pm$	1.75	&	2500	$\pm$	401	&	0.106	$\pm$	0.020	&	31.98	$\pm$	7.15	&	2364	$\pm$	125	&	5.4	\\
PN G011.3-09.4	&	PN H 2-48	&	Gaia EDR3 4078224382749921024	&	66.8	$\pm$	7.0	&	2.3	$\pm$	0.11	&	4365	$\pm$	738	&	0.049	$\pm$	0.009	&	231.20	$\pm$	32.79	&	5312	$\pm$	191	&	21.7	\\
PN G011.7-00.6	&	NGC 6567	&	Gaia EDR3 4094749870714268544	&	159.6	$\pm$	12.5	&	3.8	$\pm$	0.01	&	2725	$\pm$	260	&	0.050	$\pm$	0.005	&	202.84	$\pm$	15.89	&	3334	$\pm$	71	&	22.4	\\
PN G016.4-01.9	&	PN M 1-46	&	Gaia EDR3 4103910524954236928	&	83.5	$\pm$	9.0	&	5.8	$\pm$	1.16	&	2303	$\pm$	84	&	0.064	$\pm$	0.007	&	45.82	$\pm$	10.41	&	3262	$\pm$	176	&	41.6	\\
PN G017.6-10.2	&	PN A66 51	&	Gaia EDR3 4086643583803222400	&	29.0	$\pm$	5.8	&	31.6	$\pm$	6.31	&	1758	$\pm$	95	&	0.269	$\pm$	0.031	&	0.53	$\pm$	0.12	&	1966	$\pm$	104	&	11.8	\\
PN G025.3+40.8	&	IC 4593	&	Gaia EDR3 4457218245479455744	&	92.0	$\pm$	18.0	&	7.0	$\pm$	1.39	&	2546	$\pm$	294	&	0.086	$\pm$	0.013	&	34.86	$\pm$	9.75	&	2916	$\pm$	204	&	14.5	\\

\hline
\end{tabular}
}
\end{table}

\section{Gaia versus prior parallax measurements} \label{Gaia versus prior parallax measurements}
Surveying the literature, there were $\sim 40$ ground and space-based trigonometric PN parallaxes known prior to the Gaia period. \citet{Acker98} provided the HIPPARCOS parallax ($\pi_{\rm HIP}$) measurements for a set of 19 PNe, two of which had questionable parallaxes (SwSt\,1 \& Hu\,2-1).
In general, the parallax accuracy of this set of CSs is relatively poor since their magnitudes are close to the magnitude limit of the HIPPARCOS observatory.  Another set of 16 PN parallaxes ($\pi_{\rm USNO}$) has been measured through the US Naval Observatory (USNO) parallax program \citep{Harris07}. Twelve objects in this set had parallax errors less than $20\%$. Using the Hubble Space Telescope (HST), \citet{Benedict2009} reported highly accurate parallaxes ($\pi_{\rm HST}$) of four PNe. We derive median parallax errors of 2.5 mas, 0.42 mas, and 0.23 mas for the HIPPARCOS, USNO, and HST PN sets, respectively. The Gaia parallaxes are compared to the HST and USNO parallaxes in Figure \ref{Figure2}. In both graphs, the solid diagonal line indicates the 1:1 matches, while the dashed line illustrates the linear fitting which demonstrates a tight correlation between both the HST and USNO and the Gaia parallaxes. The HST parallaxes appear marginally smaller than the Gaia (Figure \ref{Figure2}, left panel), whereas a few USNO parallaxes are larger and others are slightly smaller than the Gaia parallaxes (Figure \ref{Figure2}, right panel). The median scaling factors $< \pi_{\rm HST}$/$\pi_{\rm Gaia} >$ and $< \pi_{\rm USNO}$/$\pi_{\rm Gaia} >$ are 0.9 and 1.0. The median errors of the HST and USNO parallaxes are $\sim 4.0$ and $\sim 8.0$ times that of Gaia (0.053 mas), respectively. In contrast to the HST and USNO results, the HIPPARCOS shows a small median scaling factor $<\pi_{\rm HIP}$/$\pi_{\rm Gaia}>$ of 0.6. Furthermore, the median error of HIPPARCOS is $\sim 40$ times that of Gaia. In this analysis, we adopt only $\pi_{Gaia}$ measurements with uncertainties less than $15\%$.

\section{Gaia distances versus other individual distances}

\begin{figure*}
	\includegraphics[width=15.0cm, height=18cm]{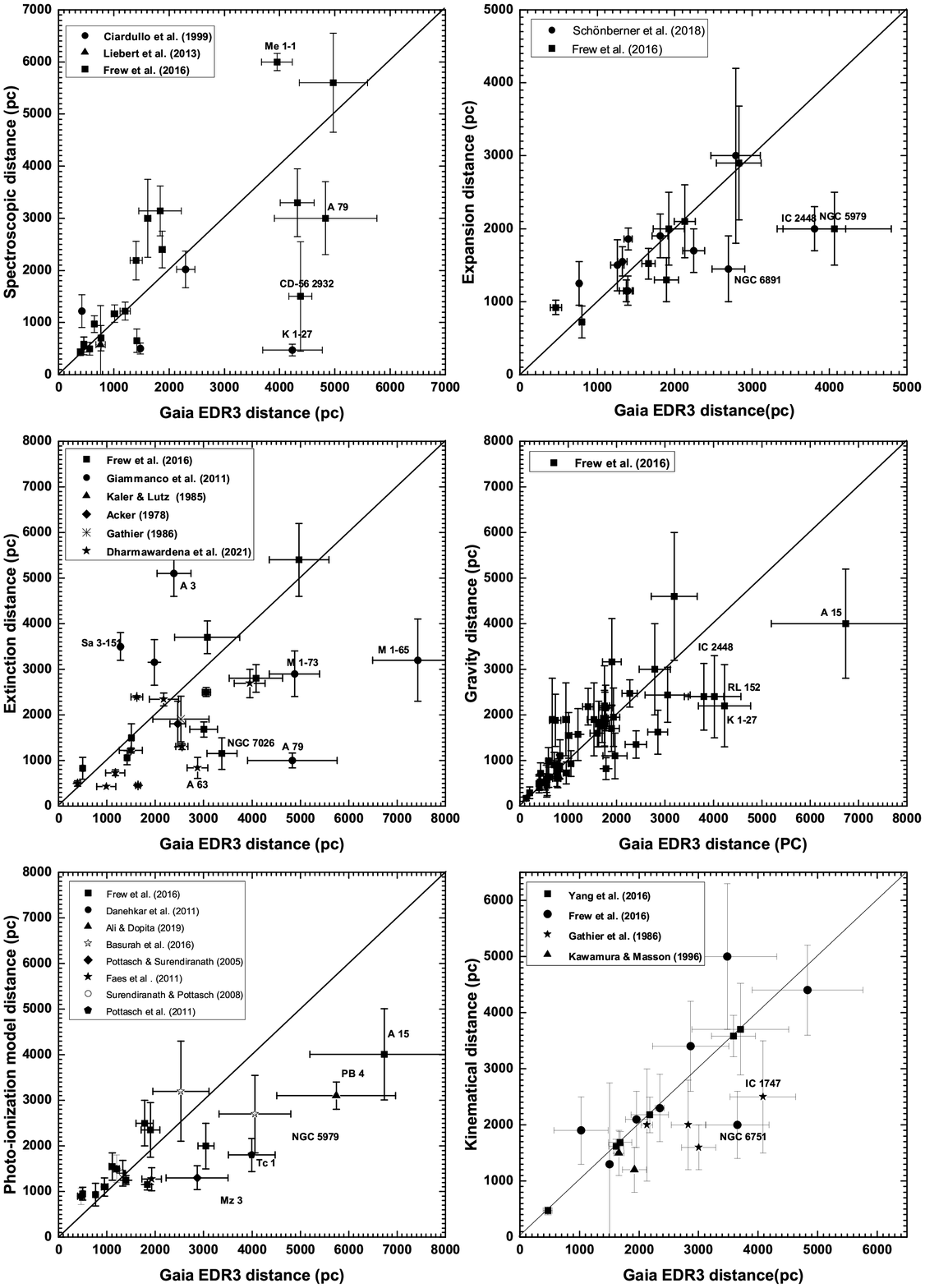}
\caption{Comparisons between the trigonometric and other individual distance methods. The Gaia versus spectroscopic distance is illustrated in the upper left panel, expansion distance in the upper right panel, extinction distance in the middle left panel, gravity distance in the middle right panel, photo-ionization model distance in the lower left panel, and kinematical distance in the lower right panel. In each graph, we distinguished only the names of outlier objects. The references for the individual distances are as follows: \citet{Ciardullo99}; \citet{Liebert13}; \citet{Frew16};
\citet{Schonberner18}; \citet{Giammanco11}; \citet{Kaler85}; \citet{Acker78}; \citet{Gathier86A}; \citet{Dharmawardena21}; \citet{Danehkar12}; \citet{Ali19};\citet{Basurah16}; \citet{Pottasch05}; \citet{Faes11}; \citet{Surendiranath08}; \citet{Pottasch11};\citet{Yang16}; \citet{Frew16}; \citet{Gathier86B}; \citet{Kawamura96}.} \label{Figure3}

\end{figure*}

As previously stated, a limited number of PN parallaxes were known prior to the Gaia mission. As a result, the previous studies on the topic of consistency between trigonometric and other individual distance methods are statistically unreliable. In AIA15, we compared directly the PN trigonometric distances to the extinction and gravity distances and indirectly, due to the small number of common objects, to the spectroscopic, expansion, photo-ionization, and kinematic distances.  The currently available number of PNe parallaxes is suitable to conduct such a study based on a statistically better basis. Moreover, to make such analyses more reliable than past ones, we adopt only Gaia distances with uncertainties less than $25\%$.  We found 23, 19, 27, 53, 20, 19 common objects between the trigonometric and spectroscopic, expansion, extinction, gravity, photo-ionization, and kinematic methods, respectively.

Figure \ref{Figure3} (top left panel) shows the spectroscopic against trigonometric distances. The comparison shows that there is a discrepancy between the two techniques for determining the distances for about half the common sample. The spectroscopic approach overestimates the distances of a few objects, e.g. Me 1-1, while underestimating the distances of others, e.g. K 1-27 and A79. The median scaling ratio $<spectroscopic/trigonometric>$ indicates that the spectroscopic technique marginally overestimates the trigonometric distances.

The comparison between the expansion and trigonometric distances is present in Figure \ref{Figure3} (top right panel). Except for a few objects (e.g. IC 2448, NGC 5979, and NGC 6891), the distances of most PNe are marginally consistent. The derived median scaling ratio $<expansion/trigonometric>$ is 0.99.  The linear regression shows a strong correlation between both distance methods (see, Table \ref{Table3}).

In Figure \ref{Figure3} (middle left panel), we compare the extinction with trigonometric distances. The inconsistency between the two methods is obvious. The extinction method underestimates the distances of more than half the common objects. The median distance ratio is 0.75. This result differs from that mentioned in AIA15, which shows a median ratio of 1.0. In support of this result, \citet{Dharmawardena21} have derived the extinction distances for a collection of 17 PNe by applying three distinct 3D extinction mapping methods and comparing them to the Gaia DR2.  We estimate the median distance ratios for the proposed three methods, which are 0.63, 1.1, and 0.83 with an average of 0.85.
Our analysis here is based on a larger statistical sample than that given by AIA15 and \citet{Dharmawardena21}. Because the usage of this method is limited to objects near the Galactic plane, we exclude objects at high Galactic latitudes from our analysis, to avoid underestimating their distances. The linear fitting also reveals a weak correlation between the two methods ($r = 0.38$). Therefore, caution should be taken when using this method to determine the PN distances.

The comparison between the gravity and trigonometric methods is plotted in Figure \ref{Figure3} (middle right panel). The plot shows the majority of objects have inconsistent distances, e.g. A15, K 1-27, RL 152, and IC 2448. In comparison to the trigonometric technique, the result reveals a slight overestimate of the gravity method, whereas the median distance ratio is 1.06. This result differs from the prior values of 0.65, 0.60, 0.77, and 0.81 provided by \citet{Napiwotzki01}, \citet{Jacoby95}, \citet{Harris07}, and AIA15, respectively. \citet{Smith15} suggested that the gravity method is distance-dependent, meaning that it overestimates the distance to nearby objects while underestimating the distance to remote objects. Figure \ref{Figure3} (middle right panel) confirms this result, where the gravity method gives overestimation and underestimation for the PNe of distances less than and greater than 3000 pc, respectively. It is significant to denote here that all gravity distances used in this comparison are compiled from FBP16, where they established an internally consistent data set using appropriate and modified parameters better than prior ones found in the literature.

The photo-ionization distances are compared with the trigonometric distances in Figure \ref{Figure3} (lower left panel).  The figure shows the photo-ionization method underestimates the distances of objects at a distance roughly less than 2000 pc while overestimating the distances of objects farther than 2000 pc. In general, there is a clear inconsistency between the distances determined by both methods.

The distances computed by the kinematical method are compared to those derived by the trigonometric method in Figure \ref{Figure3} (lower right panel). The figure shows that the consistency between both methods is higher than the other methods discussed above. The mean scaling ratio and the correlation coefficient between both methods indicate a good match.

Summarizing the previous results, there is inconsistency between the trigonometric and the spectroscopic, extinction, gravity, and photo-ionization model distance methods. The expansion and kinematical methods show moderate consistency with the trigonometric method. Table \ref{Table3} compares the mean and median scaling ratios, correlation coefficients ($r$), and probability p-values (for a null correlation) of the individual methods to the trigonometric method. The evaluated p-values for all individual methods provide strong evidence against null correlations.

\begin{table}
\centering
\caption{The mean and median scaling ratios, correlation coefficient ($r$), and the probability p-value between the individual methods and the trigonometric method.} \label{Table3}
\scalebox{0.95}{
\begin{tabular}{lcccccc}
\hline
Method	&	Spectroscopic	&	Expansion	&	Extinction	&	Gravity	&	photo-ionization	&	Kinematic	\\
\hline
$\#$ Objects	&	23	&	19	&	27	&	53	&	20	&	19	\\
Mean	&	1.10	&	1.00	&	0.92	&	1.14	&	1.24	&	0.88	\\
Median	&	1.10	&	0.99	&	0.75	&	1.06	&	1.10	&	0.91	\\
$r$	&	0.78	&	0.83	&	0.46	&	0.84	&	0.53	&	0.74	\\
P-value	&	1.02E-05	&	1.35E-05	&	1.45E-02	&	0.00E+00	&	1.71E-02	&	3.01E-04	\\
\hline
\end{tabular}
}
\end{table}

\begin{figure}
	\includegraphics[width=12.5cm, height=8cm]{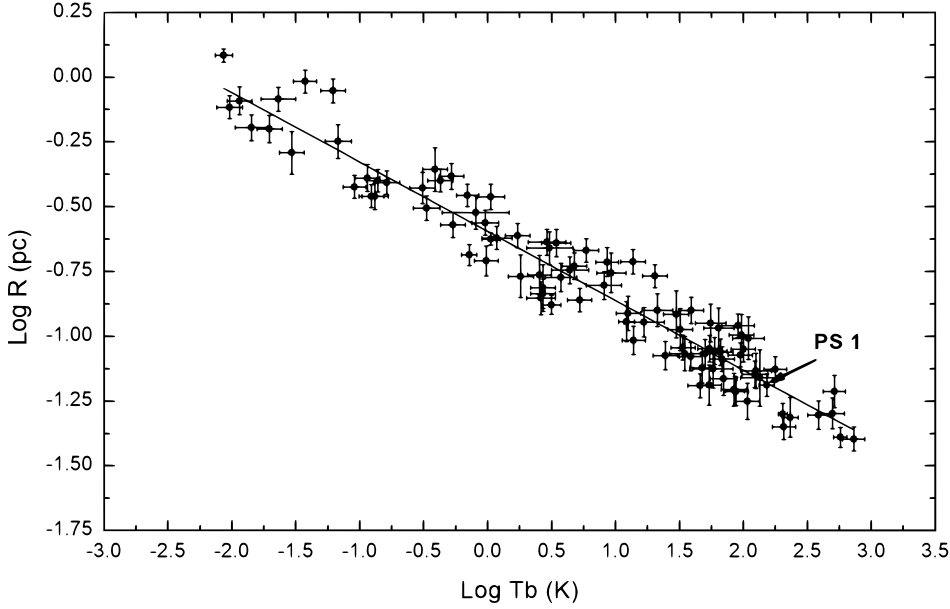}
\caption{The Tb-R relationship, based on the Gaia EDR3 calibrating sample. The black arrow shows the location of the 'PS\,1' nebula on the Tb-R relationship.}  \label{Figure4}
\end{figure}

\section{The new distance scale}
\subsection{$Tb-R$ relationship} \label{Tb-R relationship}
The $Tb-R$ relationship was first applied by \citet{vandeSteene95} and then used by others as a tool for measuring the PN statistical distance. Here, we re-construct this relationship, where the estimation of the $Tb$ and $R$ parameters is based on a precise Gaia distance sample. The $Tb$ and $R$ are estimated using equations 2 and 6 in AIA15. Figure \ref{Figure4} shows a tight anti-correlation ($r = -0.97$) between both parameters. The solid line represents the linear regression of the data points. The position of the distant nebula PS1 is indicated by a black arrow in the lower right side of Figure \ref{Figure4}. It is evident that the linear fitting does not rely on this distant nebula. From the linear fitting, we derive the equation of the new distance scale:  It is evident that the linear fitting does not rely on this distant nebula.

\begin{figure}
\includegraphics[width=12.5cm, height=8cm]{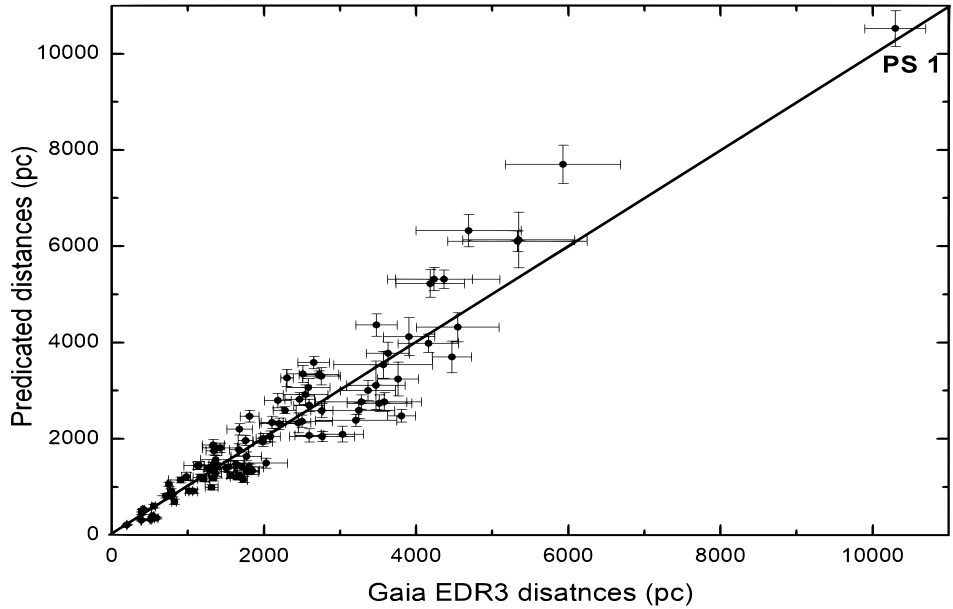}
\caption{Predicted versus calibrating distances. The solid
line indicates the 1:1 matches.} \label{Figure5}
\end{figure}

$\log (D) = 1.3817 -0.465 \log\theta -0.268 \log F_{5GHz}$

where $\theta$, $F_{5GHz}$, and $D$ are as previously defined. The uncertainty of $Tb$ and $R$ is calculated by propagating the error in the angular radius, radio flux and PN distance. To measure the quality of this approach, we compare the calibration distance ($D_{Gaia}$) with the predicted distance ($D$) in Figure \ref{Figure5}. The predicted distance of the distant "PS1" nebula according to the new scale is $10521\pm373$, which is $2.1\%$ more than the nebula's adopted distance. In general, the new distance scale slightly overestimates the PN with distances of $4.5 – 6.0$ kpc. Therefore, we should take caution about the distances of objects larger than 4500 pc.  The mean and median absolute distance residuals ($(D_G-D)/D_G)$ are respectively $18.0\%$ and $17.0\%$. This result implies that the mean error in the predicted distances is $\sim 18.0\%$, indicating that the accuracy of the new scale is better than the prior distance scales. The dispersion of the distance scale reported by \citet{Stanghellini08}, e.g., is greater than 30\%. AIA15 obtained an accuracy of $28.7\%$ for the Tb-R distance scale, while FBP16 determined distance dispersions of $28\%$ and $18\%$ for optically thick and thin PNe, respectively, with a mean accuracy of $23\%$ for the entire distance scale.


\subsection{Distance catalogue} \label{Distance catalogue}
In Table \ref{Table4}, we present a portion of the Galactic PN statistical distance catalogue which contains $\sim 1000$ PNe. The entire catalogue will be available online. Columns 1, 2-3, and 4-5 give the PN usual name, equatorial coordinates, and Galactic coordinates, respectively. Columns 6 and 7, respectively, list the adopted $F_{5GHz}$ and $\theta$ measurements and their associated errors. The predicted distances and their associated errors are given in columns 8, while in columns 9, 10, and 11 we present the distances derived by AIA15, FBP16, and \citet{Stanghellini2020}, for comparison. This catalogue will be a significant source for future PN investigations and a useful guide for the PNe of unknown and unreliable distances.

\begin{table*}
\centering
\caption{Distance catalogue.} \label{Table4}
\scalebox{0.65}{
\begin{tabular}{lcccccccccc}
\hline
PN name 	&	\multicolumn{2}{c}{Galactic coordinate}	&	\multicolumn{2}{c}{Equatorial coordinate} &	$F_{5GHz}$  &	Angular radius	&	\multicolumn{4}{c}{Distance (pc)} \\  \cline{2-3}  \cline{4-5} \cline{8-11}
            &  L  & B  & $\alpha$ & $\delta$ & (mJy)& (arcseconds) & This work & \citet{Ali15} & \citet{Frew16} & \citet{Stanghellini2020}\\
\hline
PN M  1-51    	&	20.9993	&	-1.1251	&	278.371	&	-11.124	&	319.0	$\pm$	32.0	&	6.7	$\pm$	1.3	&	2125	$\pm$	206	&	1847	$\pm$	175	&	2310	$\pm$	750	&	2270	$\pm$	740	\\																																																																																																																																																																																			
IRAS 18252-1016                    	&	21.1653	&	0.4755	&	277.006	&	-10.236	&	131.7	$\pm$	12.4	&	1.2	$\pm$	0.0	&	6042	$\pm$	153	&	4721	$\pm$	176	&				&				\\																																																																																																																																																																																			
PN M  1-63    	&	21.1704	&	-5.9834	&	282.879	&	-13.177	&	10.0	$\pm$	2.0	&	2.1	$\pm$	0.4	&	9294	$\pm$	997	&	8452	$\pm$	914	&				&				\\																																																																																																																																																																																			
IRAS 18303-1043                    	&	21.3425	&	-0.8423	&	278.277	&	-10.689	&	17.6	$\pm$	2.7	&	1.3	$\pm$	0.2	&	10079	$\pm$	791	&	8595	$\pm$	732	&				&				\\																																																																																																																																																																																			
VSP 2-19                           	&	21.6657	&	0.8109	&	276.940	&	-9.637	&	52.9	$\pm$	4.8	&	0.7	$\pm$	0.1	&	9641	$\pm$	411	&	7519	$\pm$	387	&				&				\\																																																																																																																																																																																			
IRAS 18305-1022                    	&	21.6850	&	-0.7376	&	278.343	&	-10.337	&	36.5	$\pm$	6.0	&	3.8	$\pm$	0.0	&	4921	$\pm$	218	&	4466	$\pm$	250	&				&				\\																																																																																																																																																																																			
PN M 3-55	&	21.7431	&	-0.6726	&	278.312	&	-10.255	&	19.0	$\pm$	3.8	&	8.8	$\pm$	1.8	&	3984	$\pm$	427	&	3976	$\pm$	392	&	5860	$\pm$	2460	&	6270	$\pm$	2630	\\																																																																																																																																																																																			
PN M  3-28    	&	21.8201	&	-0.4778	&	278.172	&	-10.097	&	33.4	$\pm$	3.9	&	4.7	$\pm$	0.2	&	4571	$\pm$	162	&	4236	$\pm$	176	&	3610	$\pm$	1140	&	3860	$\pm$	1220	\\																																																																																																																																																																																			
IRAS 18316-1010                    	&	21.9972	&	-0.8837	&	278.621	&	-10.127	&	14.2	$\pm$	1.7	&	1.2	$\pm$	0.1	&	10694	$\pm$	557	&	9198	$\pm$	535	&				&				\\																																																																																																																																																																																			
PN M  1-58    	&	22.0710	&	-3.1853	&	280.737	&	-11.115	&	60.0	$\pm$	25.0	&	3.2	$\pm$	0.6	&	4687	$\pm$	680	&	4106	$\pm$	701	&				&				\\																																																																																																																																																																																			
																																																																																																																																																																																																																		\hline
\end{tabular}
}
\end{table*}

Examining the prior PN distance scales reveals they are graded as long or short depending on whether they  overestimate or underestimate the PN distances. Following \citet{Phillips02}, we calculate the correlation coefficient and the relative scale ratio factor ($\kappa$) between the present distance scale and some commonly used distance scales. The result is given in Table \ref{Table5}. In general, the present distance scale is compatible with others within the error range. Nonetheless, it is longer than \citet{Cahn92}, \citet{vandeSteene95}, AIA15, \citet{Stanghellini2020} and  shorter than \citet{Zhang95} and FBP16. The results of AIA15 and FBP16 presented in Table \ref{Table5} reflect the previous results raised in Section 2 regarding the calibrator sample of both scales.

\begin{table}
\centering \caption{Scaling factors for 7 different distance scales
in the literature.} \label{Table5}
\scalebox{1.25}{
\begin{tabular}{lccc}
     \hline
Statistical scale & $\kappa $   &  $r$   & N \\
\hline

\citet{Cahn92}  & $0.79\pm0.23$  & $0.73$ & 1008  \\
\citet{vandeSteene95} & $0.94\pm0.24$  & $0.94$  & 1008 \\
\citet{Zhang95} & $1.02\pm0.32$  & $0.91$ & 1008 \\
\citet{Stanghellini08} & $1.04\pm0.22$  & $0.95$ & 1008 \\
\citet{Ali15} & $0.95\pm0.11$  & $0.99$ & 1008  \\
\citet{Frew16} & $1.02\pm0.38$  & $0.88$ & 612 \\
\citet{Stanghellini2020} & $0.93\pm0.34$  & 0.81 & 229 \\
\hline
\end{tabular}
}
\end{table}

\section{Central stars radial velocity and variability}
The stellar radial velocity (RV) and variability that are given in Gaia EDR3 were copied from Gaia DR2. The upcoming full third release of Gaia mission is expected to witness new measurements for both parameters as well as updating existing estimations. In general, the RV of PNe is measured by the Doppler shift of their emission spectral lines \citep{Durand98}. Accurate measurement requires a high-dispersion nebular spectrum. Gaia offers yet another mechanism for measuring the RV using the Doppler shift of the central star spectral lines. As a byproduct of extracting the PN parallaxes, we detect the RV for 14 CSs.  Table \ref{Table6} compares this result with the available nebular RVs in \citet{Durand98}. The stellar RV of SaSt 2-12 is the only one that is consistent with the RV of its host nebula, whereas the other five objects exhibit discrepancies. Moreover, we detect the CS variability of  ETHOS 1, HFG1, and Hen 2-447. ETHOS 1 is a close binary CS with an orbital period of 0.535 d and an extremely large amplitude \citep{Miszalski11}. HFG1 is a close detached pre-cataclysmic binary in which the CS binary components consist of a primary O-type subdwarf and a secondary F5-K0 main-sequence star \citep{Chiotellis16}. To our knowledge, the variability of Hen 2-447 central star has been detected for the first time.

\begin{table}
\centering
\caption{PNe radial velocity from Gaia EDR3 } \label{Table6}
\scalebox{1.25}{
\begin{tabular}{llcc}
\hline

Object name	&	Gaia EDR3 designation	&	\multicolumn{2}{c}{Radial velocity (km/s)}	\\
\cline{3-4}	 \\				
			&	                    & Gaia EDR3		&		\citet{Durand98}	\\	
\hline
A 35	&	3499149202247569536	&	-40.5	$\pm$	5.8	&	-6.6	$\pm$	3.8 \\
H 3-75	&	3340384082588168960	&	6.9	$\pm$	9.0	&	22.9	$\pm$	2 \\
K 1-14	&	4556040392088375168	&	-19.1	$\pm$	2.4	&	\\		
K 1-6	&	2288467186442571008	&	-47.4	$\pm$	10.7	&	\\		
K 2-7	&	6868431267915481088	&	-17.8	$\pm$	0.4	&	\\		
LoTr 1	&	2917223705359238016	&	19.3	$\pm$	3.2	&	\\		
LoTr 5	&	3958428334589607552	&	-10.5	$\pm$	1.6	&	\\		
M 1-2	&	360112911622101120	&	-22.9	$\pm$	19.6	&	-12.1	$\pm$	2 \\
M 1-44	&	4052553745525657600	&	-4.3	$\pm$	1.1	&	-107.7	$\pm$	23.4 \\
Pe 1-11	&	4042468784326846336	&	-12.0	$\pm$	0.3	&	-130.6	$\pm$	14 \\
PHR J0701-0749	&	3052395775097859072	&	46.3	$\pm$	4.1	& \\			
PN G019.5-04.9	&	4105051680543027584	&	-23.6	$\pm$	0.7	& \\			
SaSt 2-12	&	5923760667266961408	&	-67.0	$\pm$	1.6	&	-63	$\pm$	0.5 \\
WeBo 1	&	465640807845756160	&	-11.6	$\pm$	16.2	&	 \\		

\hline
\end{tabular}
}
\end{table}

\section{Conclusions}

We established a new distance scale for PNe by re-calibrating the $Tb-R$ relationship with 96 CSs, the distances of which were estimated using Gaia EDR3 trigonometric parallaxes. The advantage of using this calibration sample is that all distances are obtained using a single approach. Moreover, all the calibrators have uncertainties in their distances less than $20\%$. As a result, we created a statistical distance catalogue for $\backsim 1000$ PNe. In addition, we investigated the consistency between the trigonometric and other individual distance methods, whereas we found that most of the PN distances obtained by these methods are incompatible with the trigonometric method. The expansion and kinematical distance methods showed better consistency than other methods. In contrast to previous results in the literature, we found the extinction method underestimates the PN distances by $\backsim 25\%$ on average. This result is relatively consistent with the recent results of \citet{Dharmawardena21}. The gravity method showed overall comparable distances to the trigonometric method but differed from the previous findings in the literature that indicated that this method underestimates the PN distances. As a byproduct of extracting the PN parallaxes from Gaia EDR3 database, we identified the radial velocity for 14 PN CSs and the variability for three PN CSs, one of which was found for the first time.

\begin{acknowledgements}
This work has made use of data from the European Space Agency (ESA) mission Gaia, processed by the Gaia Data Processing and Analysis Consortium(DPAC). This research has made use of the SIMBAD database, operated at CDS, Strasbourg, France. The authors would like to thank the anonymous referee for the valuable comments and suggestion.
\end{acknowledgements}

\bibliographystyle{raa}
\bibliography{ms2022-0044-bibtex}

\end{document}